\documentclass[graybox]{svmult}

\usepackage{mathptmx}       % selects Times Roman as basic font
\usepackage{helvet}         % selects Helvetica as sans-serif font
\usepackage{courier}        % selects Courier as typewriter font
\usepackage{type1cm}        % activate if the above 3 fonts are
\usepackage{makeidx}         % allows index generation
\usepackage{graphicx}        % standard LaTeX graphics tool
\usepackage{multicol}        % used for the two-column index
\usepackage[bottom]{footmisc}% places footnotes at page bottom
\usepackage{url}
\makeindex             % used for the subject index
\begin{document}

\title*{Probing empirical contact networks by simulation of spreading dynamics}
\author{Petter Holme}
\institute{Institute of Innovative Research, Tokyo Institute of Technology, Tokyo, Japan, \email{holme@cns.pi.titech.ac.jp}}
\maketitle

\abstract*{Disease, opinions, ideas, gossip, etc.\ all spread on social networks. How these networks are connected (the network structure) influences the dynamics of the spreading processes. By investigating these relationships one gains understanding both of the spreading itself and the structure and function of the contact network. In this chapter, we will summarize the recent literature using simulation of spreading processes on top of empirical contact data. We will mostly focus on disease simulations on temporal proximity networks---networks recording who is close to whom, at what time---but also cover other types of networks and spreading processes. We analyze 29 empirical networks to illustrate the methods.
}

\abstract{Disease, opinions, ideas, gossip, etc.\ all spread on social networks. How these networks are connected (the network structure) influences the dynamics of the spreading processes. By investigating these relationships one gains understanding both of the spreading itself and the structure and function of the contact network. In this chapter, we will summarize the recent literature using simulation of spreading processes on top of empirical contact data. We will mostly focus on disease simulations on temporal proximity networks---networks recording who is close to whom, at what time---but also cover other types of networks and spreading processes. We analyze 29 empirical networks to illustrate the methods.}

\section{Introduction}

Spreading processes\footnote{In the social science and computer science literature these are commonly known as \textit{diffusion} processes. In this chapter, we stick to the natural science convention (reserving ``diffusion'' for processes where the total mass or amount of whatever is spreading, or diffusing, is conserved).} affect people at many levels. They are the basis of innovation processes~\cite{bettencourt} and spreading~\cite{abrahamson}, they shape our opinions~\cite{may15}, lifestyles~\cite{christakis_fowler} and they are also part of the mechanisms giving us infectious disease~\cite{andersonmay,gies,hethcote}. Even though disease spreading is a bit special and not the primary topic of this book, much of the theory of the interaction of spreading and contact structure comes from epidemiology of infectious diseases. For this reason, we will mostly use disease spreading as our model spreading dynamics and leave it to the reader to draw the analogies to other phenomena.

At the time of writing ``data science'' is a buzzword. One key idea behind it is that we can understand much of the social world around us by analyzing the data we create---be it from the location traces of our smart-phones~\cite{phones}, transportation cards~\cite{sun}, etc. A special type of such data records contacts between pairs of people. By contact we will mean any kind of binary interaction where something can spread from one person to the other. It could be being in physical proximity, so that disease could spread, or following someone on a social media channel, so that information could spread. A contact network could thus be thought of as a list of pairs of individuals, annotated with the time, type and locations of the interaction. In practice one usually do not have access to so much meta-information of the contacts. In many cases one must settle with only the time and identities of the individuals (a temporal network~\cite{holme_saramaki,holme_modern,masuda_lambiotte}), or even just the identities (a static network)~\cite{newman:book,barabasi:book}. On the other hand, several properties of the spreading is determined by the (temporal or static) network structure---the regularities making the network differing from a purely random one---and our understanding of how they shape the spreading dynamics is still incomplete.

One challenge for understanding how the structures of contact networks affect spreading is to be able to list and quantify the relevant structures. Structures are dependent, however, and this makes it a challenge even in the simplest case of static networks. Acquaintance networks are, for example thought to contain many triangles~\cite{grano}. The presence of triangles is thought to slow down spreading~\cite{volz_cluster}. However, this does not necessarily mean that human friendships slow down spreading. In a simple network model where the density of triangles (a.k.a.\ clustering coefficient) is the only structure to control, a typical network would one very densely connected core contributing with most of the triangles, not triangles distributed all over the network as empirical friendship networks have~\cite{burda}. Thus, there are other constraints, or structures, present in the real networks that could potentially also affect spreading phenomena. An alternative approach to controlling the density of triangles would be to take empirical networks as the starting point and simulating disease spreading directly on these. To monitor the effect of triangles one could manipulate the original network, for example by randomly rewiring links~\cite{milo}. Of course, one cannot isolate network structures completely---by rewiring the network, one could presumably change e.g.\ the average path length as well. However, one would do that from a realistic part of the  space of (temporal) networks. In addition, this approach gives insights about how the network itself acts as an infrastructure for the spreading process.

The problem of straightforward approaches to understanding the effects of network structure in real spreading processes, becomes more severe the more information-rich  network representation one uses. For temporal networks---where information about the time of contacts are included---this is particularly clear. It has been observed that human behavior has an intermittent, bursty behavior~\cite{barabasi:burst,goh}. Subsequently, authors noticed that fat-tailed interevent time distributions---the hallmark of bursty activity---slows down spreading~\cite{Karsai2011,vazquez}. At the same time, other authors observed that simulated disease spreading was slowed down by randomizing the timing of contacts in some kinds of contact networks~\cite{50185}. There must thus be other temporal structures also controlling disease spreading. In this work, we explore methods to understand the relationship between network structure and spreading dynamics that take empirical networks, rather than network models, as starting point.

In the remainder of this chapter, we will go through some of the empirical networks authors have used, typical models of spreading processes for the purpose, randomization methods, and similar techniques. Finally, will discuss future prospects and relation to other methods.

\section{Networks}

In this section, we will discuss the empirical data available at the moment and some technical issues related to how to represent it mathematically or computationally.

\subsection{Data sources}

\begin{table}
\caption{\label{tab:data}Basic statistics of the empirical temporal networks. $N$ is the number of individuals; $C$ is the number of contacts; $T$ is the total sampling time; $\Delta t$ is the time resolution of the data set and $M$ is the number of links in the projected static networks. One data set (\textit{Romania}) was coarse-grained from second to minute resolution. (We consider a pair with at least one contact (in the raw data) within a minute a contact.)}
\begin{tabular}{r|rrrrrr}
Data set & $N$ & $C$ & $T$ & $\Delta t$ & $M$ & Ref.\\ \hline
\textit{Conference} & 113 & 20,818 & 2.50d & 20s & 2,196 & \cite{conference}\\
\textit{Hospital} & 75 & 32,424 & 96.5h & 20s & 1,139 & \cite{hospital} \\
\textit{Office} & 92 & 9,827 & 11.4d & 20s & 755 & \cite{office} \\
\textit{Primary School 1} & 236 & 60,623 & 8.64h & 20s & 5,901 & \cite{school} \\
\textit{Primary School 2} & 238 & 65,150 & 8.58h & 20s & 5,541 & \cite{school} \\
\textit{High School 1} & 312 & 28,780 & 4.99h & 20s & 2,242 & \cite{hschool} \\
\textit{High School 2} & 310 & 47,338 & 8.99h & 20s & 2,573 & \cite{hschool} \\
\textit{High School 3} & 303 & 40,174 & 8.99h & 20s & 2,161 & \cite{hschool} \\
\textit{High School 4} & 295 & 37,279 & 8.99h & 20s & 2,162 & \cite{hschool} \\
\textit{High School 5} & 299 & 34,937 & 8.99h & 20s & 2,075 & \cite{hschool} \\
\textit{Gallery 1} & 200 & 5,943 & 7.80h & 20s & 714 & \cite{gallery} \\
\textit{Gallery 2} & 204 & 6,709 & 8.05h & 20s & 739 & \cite{gallery} \\
\textit{Gallery 3} & 186 & 5,691 & 7.39h & 20s & 615 & \cite{gallery} \\
\textit{Gallery 4} & 211 & 7,409 & 8.01h & 20s & 563 & \cite{gallery} \\
\textit{Gallery 5} & 215 & 7,634 & 5.61h & 20s & 967 & \cite{gallery} \\
\textit{Reality} & 64 & 26,260 & 8.63h & 5s & 722 & \cite{reality} \\
\textit{Romania} & 42 & 1,748,401 & 62.8d & 1m & 256 & \cite{roman} \\
\textit{Kenya} & 52 & 2,070 & 61h & 1h & 86 & \cite{kenya} \\
\textit{Diary} & 49 & 2,143 & 418d & 1d & 345 & \cite{read} \\
\textit{Prostitution} & 16,730 & 50,632 & 6.00y & 1d & 39,044 & \cite{prostitution} \\
\textit{WiFi} & 18,719 & 9,094,619 & 83.7d & 5m & 884,800 & \cite{wifi} \\
\textit{Facebook} & 45,813 & 855,542 & 1,561d & 1s & 183,412 & \cite{mislove} \\
\textit{College} & 1,899 & 59,835 & 193d & 1s & 13,838 & \cite{college} \\
\textit{Messages} & 35,624 & 489,653 & 3,018d & 1s & 94,768 & \cite{karimi}  \\
\textit{Forum} & 7,084 & 1,429,573 & 3,141d & 1s & 138,144 & \cite{karimi} \\
\textit{Dating} & 29,341 & 529,890 & 512d & 1s & 115,684 & \cite{pok} \\
\textit{E-mail 1} & 57,194 & 444,160 & 112d & 1s & 92,442 & \cite{ebel} \\
\textit{E-mail 2} & 3,188 & 309,125 & 81d & 1s & 31,857 & \cite{eckmann} \\
\textit{E-mail 3} & 986 & 332,334 & 526d & 1s  & 16,064 & \cite{eml3} \\
\end{tabular}
\end{table}

\subsubsection{Proximity networks}

Human \textit{proximity networks} have gained much attention recently. Such data sets records when, and sometimes where, persons are in contact. At least they contain identities of the people in contact and when the contacts happen. Typically these data sets samples people connected by some circumstance---workers in the same office~\cite{taro,office}, at the same hospital~\cite{liljeros_giesecke_holme,hospital,hornbeck}, students in the same school~\cite{school,hschool,salathe,wifi}, visitors to an art gallery~\cite{gallery}, conference attendants~\cite{conference}, etc. The time limits are typically set by the experiment and in most cases running throughout one day (when the school or office is open).

Researchers have been very creative in gathering proximity networks. One common method is to equip the participants with radio-frequency identification (RFID) sensors~\cite{thebook:barrat} which records proximity of a couple of meters. Notably, the organization Sociopatterns (\url{sociopatterns.org}) provide many open access datasets. A similar performance to RFID sensors can be obtained by infrared~\cite{taro} or wireless~\cite{salathe,roman} sensors. Another type of proximity measure is to use the Bluetooth channel of smart-phones. These typically records slightly more distant contacts (the order of $10$ meters). Bluetooth-based studies typically run longer and are less constrained~\cite{reality,stopczynski2014measuring,stop}.

In addition to proximity recorded by sensors, researchers have used location information to infer who is close to whom at what time. Ref.~\cite{sun} studies people sharing the same public transport; Ref.~\cite{wifi} uses a dataset of people connected to the same WiFi router. There is also a rather large field of studying patient flow within hospital systems, e.g. Refs.~\cite{liljeros_giesecke_holme,donker,donker2012,walker} from the records of patients and healthcare workers. A contact in such networks corresponds to two persons being at the same ward at the same time.

Yet another kind of human proximity networks (perhaps different enough to constitute a stand-alone category) is sexual networks. Classic sexual network studies  do not  have the time of the contacts. The only large-scale temporal network of sexual contacts we are aware of is the prostitution data of Rocha et al.~\cite{prostitution} where contacts with sex sellers are self-reported by the sex buyers at a web community.

Finally, although this book focuses on humans, we mention that proximity networks of animals have been studied fairly well. In particular, populations of livestock have been studied either as metapopulation networks (where one farm is one node and an animal transport between two farms is a contact), or as a temporal network of individual animals where a contact represents being at the same farm at the same time. Livestock here could refer to either cattle~\cite{gates_woolhouse,scholtes_causality,valdano_poletto} or swine~\cite{konschake_components}.  In addition domesticated animals, researchers have  also studied wild animals---zebras~\cite{lahiri_berger_wolf_2007} and monkeys~\cite{capuchin} by GPS traces, ants~\cite{thebook:charbonneau} by visual observation, and birds~\cite{psorakis} from foraging records.

In the latter part of this chapter we will use some data set of this type as an example. In particular, we use several Sociopatterns data sets: \textit{Conference} (participants of a computer science conference), \textit{Hospital} (patients, doctors and nurses of a hospital),  \textit{Office} (workers at the same office), \textit{Primary} and \textit{High School} (school students), \textit{Gallery} (visitors to an art gallery), and families in rural Kenya (\textit{Kenya}). Some of these data sets covers several days, which we treat separately. We also use one Bluetooth data set sampled among college students in USA (\textit{Reality}) and a similar dataset from Romania (\textit{Romania}) sampled with WiFi technology. Finally, we use one dataset based on a diary-style survey (\textit{Diary}) and one from self-reported sexual contacts with escorts (\textit{Prostitution}). Statistics and references to these data sets can be found in Table~\ref{tab:data}.

\subsubsection{Communication networks}

Temporal networks of human communication are probably the largest class of systems modeled as temporal networks after proximity networks. One such type of data comes from call-data records of mobile phone operators~\cite{bernhardsson,Karsai2011,kovanen_motif}. These use lists who called whom, or who texted whom. Typically the data sets are restricted to one operator in one country. Another type of communication networks are e-mails sampled from the accounts of a group of people during a window of time~\cite{ebel,eckmann}. Yet another of this kind comes from messages at social media platforms such as Twitter~\cite{romero,sanli} or Internet communities~\cite{pok,karimi_ramenzoni,blasio,jacobs_thefacebook,mathiesen}. A difference to proximity networks is that links in this category are naturally directed. (Later in this chapter, when we will compare networks of this kind to proximity networks and then treat contacts as undirected.)

Below we analyze one data set of wall-posts at Facebook (\textit{Facebook}), one Facebook-like community for college students (\textit{College}), one Internet dating service (\textit{Dating}) and one film-discussion community (\textit{Forum} for posts at a discussion forum, and \textit{Messages} for direct, e-mail-like communication). We also study three data sets of e-mail communication \textit{E-mail 1}, \textit{2} and \textit{3}. A summary of these data sets and references can be found in Table~\ref{tab:data}.

\subsection{Network representations}

The basic setting we are considering is a set $V$ of $N$ \textit{nodes} (sometimes called \textit{vertices}). For most purposes of this chapter, the nodes represent individual people. In a static network, or \textit{graph} (emphasizing the mathematical representation rather than the real system) $G(V,E)$, the nodes are connected pairwise by $M$ \textit{links} (sometimes called \textit{edges}) $E$. In a temporal network the nodes are connected at specific times by $C$ \textit{contacts} (sometimes called \textit{events})---triples $(i,j,t)$ showing that $i$ interacted with $j$ at time $t$. An alternative way of thinking about how time enters networks is to consider nodes as a sequence of static graphs  $\{G_t(V_t,E_t)\}_{t=1}^T$, one for every discrete time step of the data. $T$ is called the \textit{sampling time}. This type of graph sequence is a special case of \textit{multilayer} networks~\cite{kivela_rev,boccaletti_rev}. Mathematically it is equivalent to sequences of contacts, but it does put other ideas into the mind of the user. To be specific, thinking of the system as a sequence of graphs suggests that one can first apply static network theory to each time slice individually, then aggregate these results. This could be a powerful approach in many cases, but not if the time resolution is so high that the networks are mostly very fragmented (or perhaps even empty, as the case in e.g.\ an e-mail network). Since paths in temporal networks need to follow the arrow of time, they are not transitive---the pairs $(i,i')$ and $(i',i'')$ can have contacts without $i$ being able to influence $i''$. The reason is that all contacts between $(i',i'')$ might have happen by the time the spreading has reached $i''$.

Many studies consider spreading processes in space. This is true not only for disease spreading~\cite{spatial_epi}, but the study of spreading of innovation (Ref.~\cite{mcvoy1940patterns} is an important early such reference). In principle, space can be encoded into the contacts of a network. On the other hand, in cases one are not aware of the detailed contact structure, one can resort to spatial spreading models. Spatial information can be combined with a network representation~\cite{spatial} and such mixed approaches are efficient in modeling multi-scale human mobility patterns (and thus contact patterns)~\cite{Balcan22122009}.

\section{Spreading dynamics}

In this section we, discuss models of spreading phenomena that can be simulated on empirical contact sequences. It is not a complete review of the matter, but intended to give a make the latter discussion more concrete.

\subsection{Epidemic spreading}

The framework for modeling the spread of infections in a population is well established~\cite{gies,andersonmay}. So called \textit{compartmental models} divide the population into states (classes, or compartments) with respect to the disease, and prescribe transition rules between these classes. The four most common states are: \textit{susceptible} (S, individuals that can get the disease, but not spread it), \textit{infectious} (I, who  can spread the disease), \textit{recovered} (R, who can neither get or spread the disease), and \textit{exposed} (E, who got the disease but can yet not infect others). The infection event typically happens between a susceptible and infectious individual. It is the only transition that requires two people to meet. Two canonical compartmental models are the SIR and SIS model. In SIR an S person can become I upon meeting an S, and an I will eventually become R. In SIS, I becomes S rather than R. There are some subtleties involved in how to implement the transitions. Mathematical epidemiology has traditionally implemented the transition from I (to R in the SIR model or S in the SIS model) as happening with a constant rate. In other words, all infected persons have the same chance of becoming uninfected every unit of time. This---\textit{constant infection rate} (CIR) version---leads to an exponential distribution of the infection time, which is in contrast to observations~\cite{vergu}, but simplifies the calculations. An other approach---the \textit{constant infection duration} (CID) version---is to model the duration of the infection as constant. As all infected nodes expose their neighbors the same amount of time, this simplifies some statistical analyses. It is also somewhat algorithmically more straightforward (but this should not be a ground for selecting the algorithm). In either of these cases, the SIR and SIS models have two parameter values. One controlling how easily a node gets infected. Another controlling how long the node stays infected. For the CIR version these are the infection and recovery rates. For the CID version they are the per-contact infection probability $\lambda$ and disease duration $\delta$.

The second ingredient in epidemic modeling is a model or data of the contact patterns. In most approaches this part comes from a simple model---the simplest being that everyone has the same chance of meeting everyone else at every time, but, in particular, with the advent of network epidemiology~\cite{keeling_rev}, researchers have started to study more realistic contact patterns. One approach is to construct models of human contact patterns. For example, based on the observation that sexual networks have a power-law degree distribution, researchers have studied the transmission of sexually transmitted infections on model networks with such a degree distribution~\cite{lea:sex}. Another approach for increased realism in disease spreading studies is to simulate the spreading on data sets of empirical contacts~\cite{salathe,50185,read,holme_tempdis,holme_masuda_r0}. As mentioned, this approach gives more than better  predictions---we can also use it to understand what structures of the contact sequence that is important, and why.

\subsection{Opinion and information spreading} \label{sec:infospread}

Much of the previous section is true for modeling of information and opinion spreading too. The main difference is that one cannot assume that such spreading is well-modeled by compartmental models. We have learned from studies of spreading in social media that individual behavior is very diverse and platform dependent. Not only do people have different activity levels, they could also follow completely different mechanisms~\cite{romero,DeMartino2015}. Sometimes authors make the distinction between \textit{simple} and \textit{complex contagion}~\cite{weng}. The former are all types of spreading phenomena where the spreading can be modeled as a probabilistic event when an S meets an I, independent of the rest of the system. Complex contagion, on the other hand can depend on more than a pairwise interaction: an opinion might need exposure from several different neighbors to spread from one vertex to another; a piece of information might spread slower with age; it could spread more easily between people that are similar to one another, etc.~\cite{weng}

The simplest type of complex contagion are threshold models.\footnote{An even simpler type of opinion spreading model is the \textit{voter model}~\cite{voter}. This is a simple contagion model where random nodes copy the opinion of random neighbors.} These assume that an individual adopts an idea when the exposure is over a threshold. What ``exposure'' means is not trivial. It could be the number of different persons that one hear an opinion from; it could also be the number of times one has heard the opinion~\cite{backlund_threshold}. Furthermore, for temporal networks, old exposures are not as important as more recent ones~\cite{karimi,taro_holme_threshold}. Authors have modeled this by counting exposures in a time window into the past~\cite{karimi_holme} or assigning every contact with an exponentially decreasing importance metric~\cite{taro_holme_threshold}.

\begin{figure}
  \includegraphics[width=0.9\columnwidth]{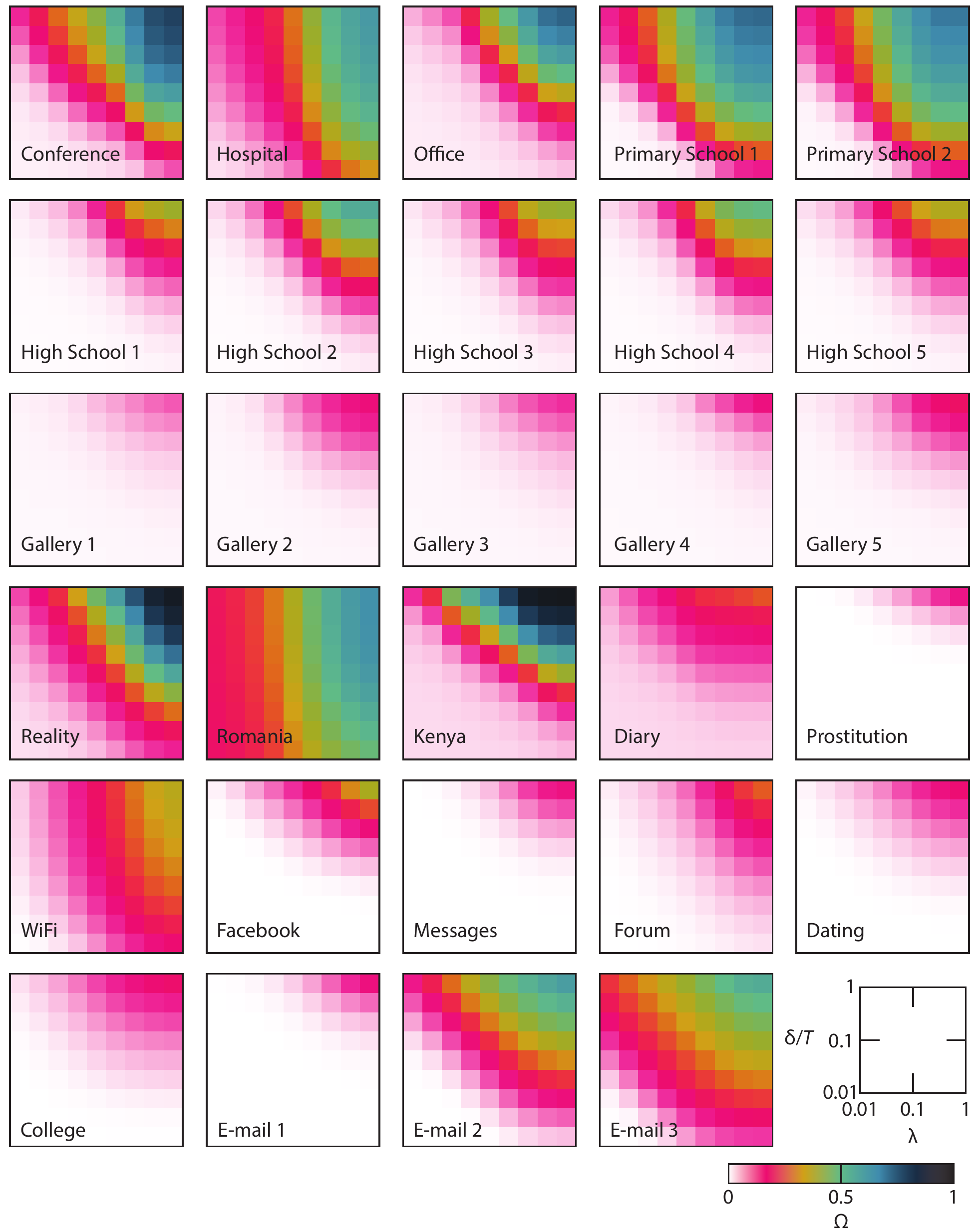}
  \caption{The average outbreak size for SIR epidemics on our data sets. The scales of the axes and colors are the same for all panels (as indicated in the legend).
}
\label{fig:ch_osze}
\end{figure}

\begin{figure}
  \includegraphics[width=0.9\columnwidth]{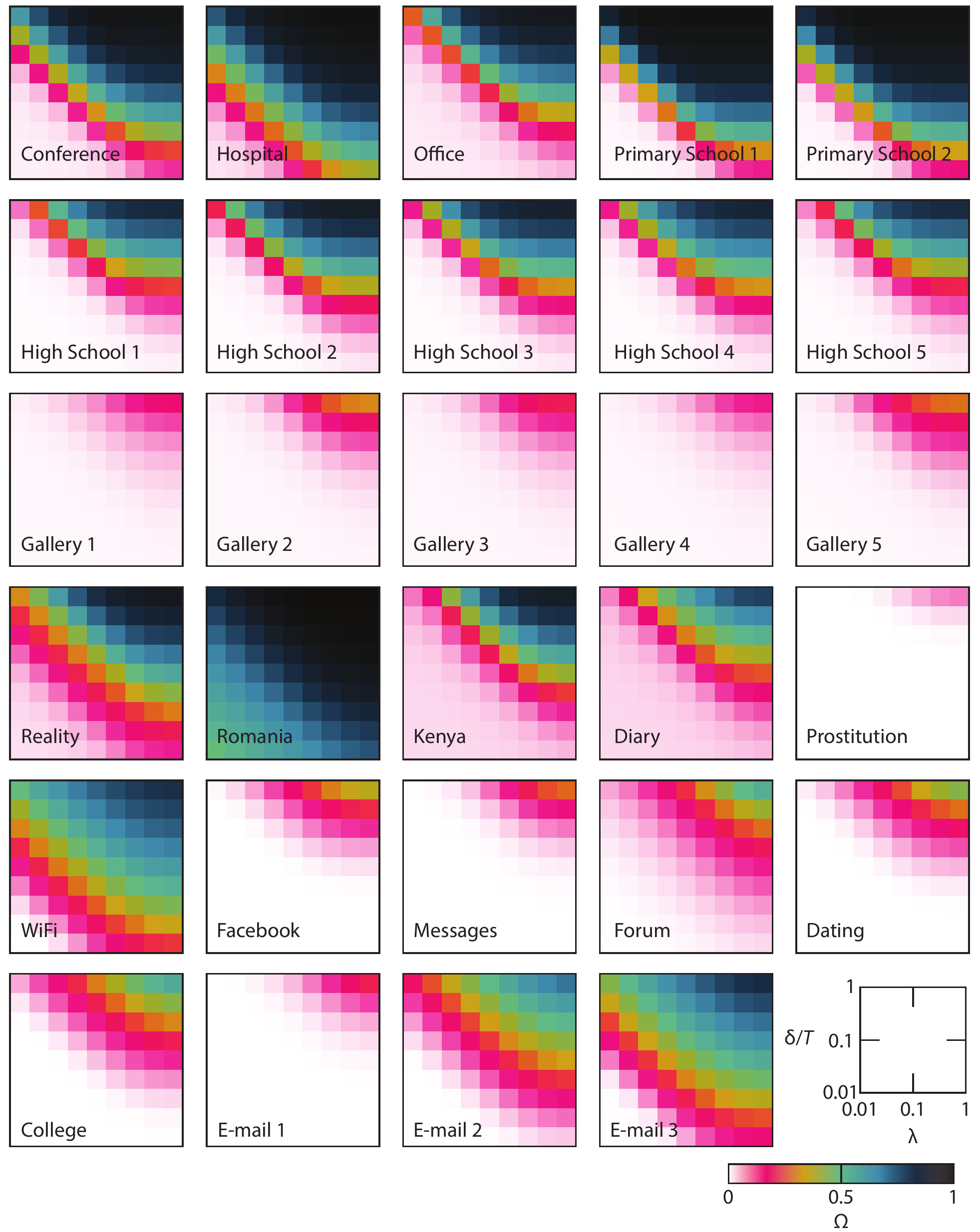}
  \caption{The average outbreak size for SIR epidemics data sets where the time stamps of contacts are replaced by random ones. Otherwise, the figure is the same as Fig.~\ref{fig:ch_osze}.
}
\label{fig:ch_osze_t}
\end{figure}

\begin{figure}
  \includegraphics[width=0.9\columnwidth]{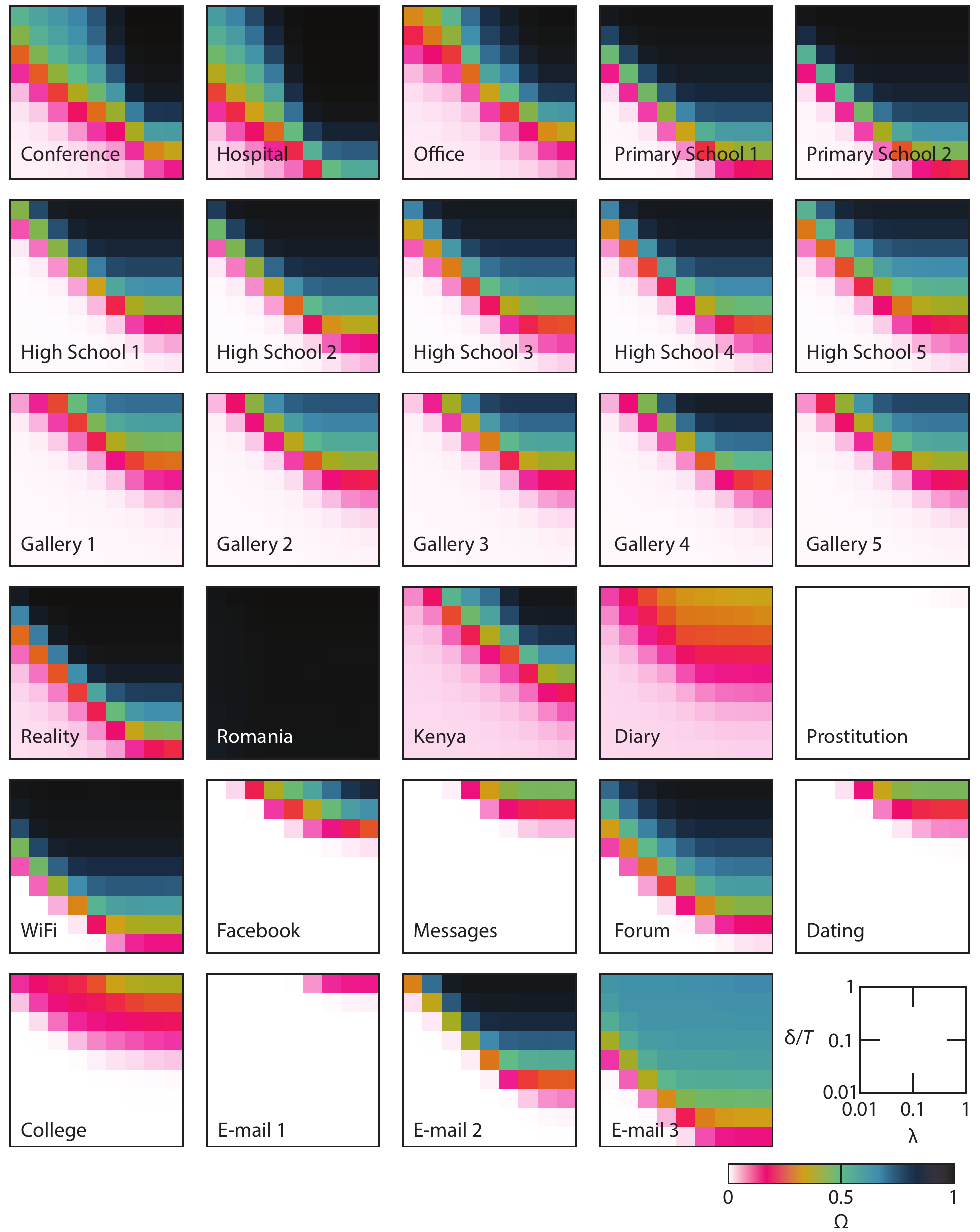}
  \caption{The average outbreak size for SIR epidemics data sets where the node identities are replaced by random ones. Otherwise, the figure is the same as Fig.~\ref{fig:ch_osze}.
}
\label{fig:ch_osze_n}
\end{figure}

\section{Null models, randomizations and positional comparisons}

So far, we have discussed the kind of datasets available and different dynamic models of spreading phenomena. While running such simulations on the raw contact data can be interesting in its own right, it can be hard to generalize the results. As mentioned in the Introductions, one option is to compare the results to those expected from models. This approach has been a fruitful way for static networks but is challenging for more information-rich representations of the contact patterns. One reason is that it is hard to even name a reasonably complete set of simple structures  in temporal networks (it is of course even harder to control them in a way such that the results are easy to interpret). An alternative approach is to draw the conclusions from comparisons. One way is to use randomize some aspect of the real data and thereby destroy some particular structure. By comparing the spreading on the original and randomized networks, one can draw conclusions about the effects of the randomized structures. By successively randomizing less and less one can, in principle, home into the important structures. One may argue that this approach is only replacing the problem of listing fundamental structures, by the problem of listing structures to randomize. However, one is certain that going from the original data to the fully randomized data, one has removed all structure there is, and thus all structure that can play a role in the spreading (even though this procedure should be coarser than ideal).

Refs.~\cite{holme_saramaki,holme_coll} presents several methods of randomization. In this chapter, we will exemplify with two: \textit{Random times} (RT) and \textit{Random links} (RL). For RT one replaces the timestamps of contacts with random times in the interval $[0,T)$, thus destroying several types of temporal structures including effects of: order of events, periodic changes in the overall activity, the turnover of individuals, etc. RT is thus a quite pervasive type of randomization, only conserving the number of contacts and the static network structure. One can regard RL as a topological counterpart to RT. For RT one replaces link by a link between two random nodes in the network. Thus one destroys all the topological structure, including the degree distribution.\footnote{Effectively one replaces the degree sequence by one drawn from a binomial distribution. For many applications one is rather interested in the topological structure other than the degree distribution, and would rather conserve the degree of the nodes~\cite{milo}, but to be able to compare the topological randomization to the temporal one, we use this definition.} With the results for the original and randomized networks at hand one can see how the destroyed structure affects the spreading. This effect would typically depend on the parameter values of the spreading dynamics. To get interpretable results one typically average over many randomized data sets. The good news is that temporal networks are typically ``self-averaging'' in the sense that fluctuations decrease with systems size. To move further into describing how the contact structure affects the spreading one can also include measurements of (temporal or static network structure). For example, Ref.~\cite{holme_masuda_r0} compares the discrepancies between two estimates of disease severity for different contact data sets. They correlate the discrepancies with measures---network descriptors---like the node and link burstiness~\cite{goh}, the fraction of nodes and links present throughout the sampling time, etc. From this analysis they can conclude that some types of discrepancies are more related to temporal structures, other to topological structures.

In addition to randomizing away structure, one can learn about the structure of the contact network by comparing nodes and links of within the same network. One can for example compare spreading starting at different nodes and compare the average outbreak size, time to peak prevalence or time to extinction~\cite{holme_pcb,rocha_masuda}. Another approach would be to eliminate single nodes and links and study the changes of the mentioned quantities~\cite{bramson}. 

\section{Example: SIR model on empirical networks}

In this section, we will present an analysis along the lines outlined above for the $29$ contact networks of Tab.~\ref{tab:data}. In Fig.~\ref{fig:ch_osze}, we show the average outbreak size $\Omega$ (the fraction of recovered nodes at the end of the outbreak) in an SIR simulation. We use the CID version, so the two parameter values are the per-contact transmission probability $\lambda$ and the disease duration $\delta$. The infection is started by one randomly chosen node at a random time between $0$ and $T$. All data points are averaged over at least $10^3$ outbreak runs per networks. In Figs.~\ref{fig:ch_osze_t} and \ref{fig:ch_osze_n} we show plots of $\Omega$ as a function of $\lambda$ and $\delta$ for the RT and RL randomizations, respectively. For these figures, we also average each value over $100$ randomizations.

A first thing to notice in Fig.~\ref{fig:ch_osze} is that $\Omega$ is increasing with both $\lambda$ and $\delta$. For some networks, $\Omega$ reaches its maximal value $1$, but for most it does not. For the \textit{Gallery} data, \textit{Prostitution} and the social media networks (\textit{Facebook}, \textit{Messages}, \textit{Forum}, \textit{Dating}, and \textit{College}) there is a big overturn of agents---the individuals that are there in the beginning are not there in the end. (This is easy to imagine for the \textit{Gallery} networks as a visitor to an art gallery would stay for a limited amount of time.) A low maximal $\Omega$ can only most easily be explained by that the turnover of agents breaks many time-respecting paths so that one individual cannot infect so many others. Another explanation would be that most contacts happen in the beginning, so that by chance the network is very fragmented by the time a typical first infection event happens. Some of the networks are so dense that even for the lowest parameter values ($\lambda=\delta/T=0.01$) $\Omega$ is quite large. The most conspicuous example is perhaps \textit{Romania} where both the minimum and maximum $\Omega$ value is intermediate. At this point, it is worth noting that $\lambda$ (unlike $\delta$) should not be understood as a parameter that is unique for one disease. It must be defined in combination with the network representation---a more restrictive definition of a contact would correspond to a larger $\lambda$ value~\cite{stop}. Finally, we note that the data sets that come from the same setup (the five \textit{High School} and \textit{Gallery} networks looks like each other---an indication that the used methods are not sensitive occasional misinformation).

Next, we turn to the data with randomized time stamps. $\Omega$ as a function of $\lambda$ and $\delta$ is displayed in Fig.~\ref{fig:ch_osze_t}. The effect of the randomization is different for different data sets. For several data sets $\Omega$ increases, at least the maximal $\Omega$, or sometimes $\Omega$ throughout the parameter space. The main exception is \textit{Prostitution} where the maximal $\Omega$ decreases upon randomization. Some other data sets---the \textit{Gallery} data and \textit{E-mail 1} do not change. From this we understand that given the underlying contact network, and the number of contacts between each pair of nodes, the timing of the nodes can both speed up and slow down the disease spreading. Since bursty behavior is known to slow down spreading~\cite{vazquez,Karsai2011} and the RT randomization makes contacts less bursty, we can understand that other temporal factors also determine the speed and scope of the spreading. We can also notice that the effects of RT randomization is largest for intermediate values of $\Omega$ (of Fig.~\ref{fig:ch_osze}).

The outbreak sizes corresponding to Fig.~\ref{fig:ch_osze} for topologically randomized datasets is shown in Fig.~\ref{fig:ch_osze_n}. The pattern from the RT plots of Fig.~\ref{fig:ch_osze_t} remains, and is somewhat accentuated. For the RL randomized plots, $\Omega$ reaches close to its maximum value $\Omega=1$ for most of the data sets (including \textit{Gallery}, where this was not true for the RT randomization). For some other datasets---\textit{Prostitution}, \textit{Diary}, \textit{E-mail 1} and \textit{E-mail 3}---the maximal $\Omega$ decreases going from RT to RL. In summary, most datasets we test have investigated have both temporal and topological structures that decreases the outbreak sizes. Why the opposite occasionally happens is an interesting and---at the moment of writing---not fully resolved problem. One observation is that all such data sets are fairy sparse in the sense that $C/M$ is small, but other structures can separate these data sets from others even more clearly. One such structure is the average duration of a link~\cite{holme_liljeros}. Other quantities that describe the long-term evolution of the system could also work well.

\section{Discussion}

In this chapter, we have pointed at some ways one can analyze empirical datasets of human interaction by simulating spreading phenomena on top of them. We have discussed data sources, network representations and some of the analysis techniques including some of the many randomization-based null-models available. The type of analysis we have outlined is not from a pipeline to handle contact data. Indeed, the analysis of temporal networks is not so developed or systematic as the ones for static networks. To understand how contact networks works, at the moment, one need to use different approaches at once (what we sketched in this chapter is one of them). Several papers that have used the same approach as this chapter do not stop after comparing the original and randomized networks, they continue to try to identify lower-level structures. For example Refs.~\cite{holme_masuda_r0,holme_tempdis} characterizes the differences between SIR spreading on the original and randomized data, then performs a regression analysis to find which low-level structural descriptor that has the highest explanatory power.

For the future, we hope it will be possible to formalize the ideas in the chapter to a more programmatic approach. In particular, it is probably possible to construct a flow-chart how to perform successive randomizations to identify the important structures for spreading on a particular data set. This would need to solve the question about how to randomize away arbitrary structures (or at least all structures that are easy to understand conceptually, thus contributing to our understanding of network dynamics). For opinion spreading problems, a major challenge is to find an appropriate microscopic model of the spreading~\cite{weng}.

\bibliographystyle{spphys}
\bibliography{probing}

\end{document}